\documentclass{article}
\usepackage{amsmath}
\usepackage{cite}
\usepackage{amssymb}
\usepackage{hyperref}
\allowdisplaybreaks
\begin{document}

\title{A first-class approach of higher derivative Maxwell-Chern-Simons
Proca model}
\author{ S. C. Sararu\thanks{
e-mail address: scsararu@central.ucv.ro} \\
Department of Physics, University of Craiova\\
13 Al. I. Cuza Str., Craiova 200585, Romania}
\date{}
\maketitle

\begin{abstract}
The equivalence between a higher derivative extension of Maxwell-Chern-Simons Proca model and some gauge invariant theories from the point of view of the Hamiltonian path integral quantization in the framework of gauge-unfixing approach is investigated. The Hamiltonian path integrals of the first-class systems take manifestly Lorentz-covariant forms.
\end{abstract}

\section{Introduction}

The quantization of a second-class constrained system can be achieved by the
reformulation of the original theory as a first-class one and then
quantizing the resulting first-class theory. This quantization procedure was
applied to various models \cite{stuckn,vytgen,kor,stuck5,stuck3,biz2formx,biz2formPRD,biz2formPRD1,stuck2,stuck4,vyt3,stuck6,stuck1,IJMPAnoi,EMC,io1,io2,io3}
using a variety of methods to replace the original second-class model to an
equivalent model in which only first-class constraints appear. The
conversion of the original second-class system into an equivalent gauge invariant theory can be accomplished without enlarging the phase
space, starting from the possibility of interpreting a second-class
constraints set as resulting from a gauge-fixing procedure of a first-class
constraints one and "undo" gauge-fixing \cite{jap,mitra,marc,vyt0,bizNPB}. The
gauge-unfixing method relies on separating the second-class constraints into
two subsets, one of them being first-class and the other one providing some
canonical gauge conditions for the first-class subset. Starting from the
canonical Hamiltonian of the original second-class system, we construct a
first-class Hamiltonian with respect to the first-class subset through an
operator that projects any smooth function defined on the phase space into
an application that is in strong involution with the first-class subset. Another method to construct the equivalent first-class theory relies on an
appropriate extension of the original phase space through the introduction
of some new variables. The first-class constraints set and the first-class
Hamiltonian are constructed as power series in the new variables \cite{FS,bfgen,bfgen1,battyut}. Various aspects of the equivalence \cite{DJ1} between self-dual model \cite%
{TPN} and Maxwell-Chern-Simons (MCS) theory \cite{DJT1,DJT11} have been studied using one
of the two methods mentioned in the above \cite{io2,BRR1,BRR11,BR1}. A generalization of the Proca action for a
massive vector field with derivative self-interactions in $D=4$ has been
constructed in \cite{LH}. In \cite%
{DJ2,SN,AD1,AD2,Baz} one finds higher derivative extensions that involve the Maxwell and/or
Chern-Simons (CS) terms \cite{DJ1,TPN,DJT1}. The Lagrangian of such model is the sum of Maxwell, CS and higher derivative extensions of these terms. The generalized MCS-Podolsky model \cite{POD,Baz} is a such theory and was introduced in order to smooth ultraviolet singularities.
Starting from the observation that the study of Einstein-Chern-Simons Proca massive gravity (ECSPMG) (the Lagrangian of ECSPMG is the sum of Einstein, (third derivative order) CS and Proca-like mass terms) \cite{PPT} is often accompanied \cite{DT,AD1,AD2} by the analysis of the MCS-Proca model (a non-higher derivative model) \cite{SN,BCSx,DT,Baz,BK}, we consider a model described by Lagrangian action containing the Maxwell term, a higher derivative extension of the CS topological invariant \cite{DJ2} and Proca mass term
\begin{equation}
S=\int d^{3}x\left[ -\frac{a}{4}\partial _{\lbrack \mu }A_{\nu ]}\partial
^{\lbrack \mu }A^{\nu ]}+\frac{1}{2b}\varepsilon _{\mu \nu \rho }\left(\partial _{\lambda }\partial ^{\lambda }A^{\mu }\right) \partial ^{\nu
}A^{\rho }
-\frac{m^{2}}{2}A_{\mu }A^{\mu }\right].  \label{y1}
\end{equation}
and we investigate from the point of view of the Hamiltonian path integral quantization using the gauge-unfixing (GU) approach the previous higher derivative extension of the MCS-Proca model. The choice of the extended MCS-Proca (MECS-Proca) model will become more transparent in the subsection \ref{sec3.1} where we will find that between the extended MCS-Proca (MECS-Proca) model and ECSPMG theory they are same similarities regarding to the number of physical degrees of freedom and the presence of ghosts and tachyon excitations. In order to construct an equivalent first-class system starting from the
MECS-Proca model in the framework of the GU approach, we need to know the
structure of the constraints set of the model. As the second term in the
action (\ref{y1}) contains higher derivative terms $\left\{ \partial
_{\lambda }\partial ^{\lambda }A_{\mu }\right\} $, the canonical analysis
will be done by a variant of Ostrogradsky method \cite{O1,GLT,GT,N1,K1,R1}
developed in Ref. \cite{BMP}, based on an equivalent first order formalism \cite{P1,P2} and applied to a number of particle and field theoretic models \cite%
{BMP,MP,Paul,BMP1}. The Hamiltonian analysis of a higher derivative extension of a theory displays a constraints set with a more complicated structure than the constraints set of the usual theory (where Lagrangian is function of the fields and their first derivatives only). The separation of a second-class constraints set with a complicated structure in two subsets (one of them being first-class and the other one providing some canonical gauge conditions for the first-class subset) is an intricate issue. In general, in the structure of the constraints set of the higher derivative extension we find a reminiscence of the structure of the constraints set of the usual theory. In order to do more transparent the approach of the MECS-Proca model, initially we consider the MCS-Proca model and we apply the quantization procedure mentioned in the above. Next, we focus on the Hamiltonian analysis of the MECS-Proca model and the construction of the equivalent first-class system using gauge-unfixing method. Then, we construct the Hamiltonian path integral of the equivalent first-class system.
After integrating out the auxiliary fields and performing some field redefinitions, we discover the manifestly Lorentz covariant path integral corresponding to the Lagrangian formulation of the first-class system, which
reduce to the Lagrangian path integral for St\"{u}ckelberg coupling between a scalar field and a $1$-form or to the Lagrangian path integral for two kinds of $1$-forms with CS coupling.

The paper is organized in four sections. In section \ref{sec2}, starting
from MCS-Proca model we construct an equivalent first-class model using
gauge-unfixing method and meanwhile we obtain the path integral
corresponding to the first-class system associated with this model. Section \ref{sec3} contains the main results of the present paper. Firstly, we perform Hamiltonian analysis and study the excitations and mass counts of the MECS-Proca model. Secondly,  we exemplify in detail the gauge-unfixing method on
MECS-Proca model and then we construct the path integral of the
equivalent first-class system associated with this second-class theory.
Section \ref{sec4} ends the paper with the main conclusions.

\section{The MCS-Proca model\label{sec2}}

The MCS-Proca model is described by the Lagrangian action \cite{SN,BCSx,DT,Baz,BK}
\begin{equation}
S=\int d^{3}x\left(-\frac{a}{4}\partial _{\lbrack \mu }A_{\nu ]}\partial
^{\lbrack \mu }A^{\nu ]}-b\varepsilon _{\mu \nu \rho }A^{\mu }\partial ^{\nu
}A^{\rho }-\frac{m^{2}}{2}A_{\mu }A^{\mu }\right) ,  \label{actmcs}
\end{equation}
where $a$ and $b$ are some real constants. We work with the Minkowski metric
tensor of `mostly minus' signature $\sigma _{\mu \nu }=diag(+--)$. The canonical analysis \cite{D1,D2} of the model described by
the Lagrangian action (\ref{actmcs}) displays the second-class constraints
(scc)
\begin{eqnarray}
\chi ^{( 1) } &\equiv& p^{0}\approx 0,  \label{pgei} \\
\chi ^{( 2) } &\equiv& \partial _{i}p^{i}-b\varepsilon
^{0ij}\partial _{i}A_{j}-m^{2}A^{0}\approx 0,  \label{pcei}
\end{eqnarray}%
and the canonical Hamiltonian
\begin{eqnarray}\label{phamcan}
H_{c}&=&\int d^{2}x\left( -\frac{1}{2a}p_{i}p^{i}-A_{0}\partial _{i}p
^{i}+\frac{a}{4}\partial _{\lbrack i}A_{j]}\partial ^{\lbrack
i}A^{j]}\right.\nonumber\\
&&\left.+b\varepsilon _{0ij}A^{0}\partial ^{i}A^{j}+\frac{b}{a}\varepsilon
_{0ij}A^{i}p^{j}-\frac{b^{2}}{2a}A_{i}A^{i}+\frac{m^{2}}{2}A_{\mu }A^{\mu
}\right),
\end{eqnarray}
where $p^{\mu }$ are the canonical momenta conjugated with the fields $%
A_{\mu }$. The number of physical degrees of freedom \cite{marc} of the
original system is equal to%
\begin{eqnarray}
\mathcal{N}_{O} &=&\left( 6\ \rm{canonical\ variables}-2\ \rm{scc}\right) /2  \nonumber \\
&=&2.  \label{doforig}
\end{eqnarray}
The same result, with respect to the number of degrees of freedom, is obtained
in Refs. \cite{BK,DT}. Moreover, in Refs. \cite{BK,DT} it is shown that the MCS-Proca model
describes a topological mass mix with two massive degrees of freedom, with
masses $\sqrt{b^{2}+m^{2}}\pm \vert b\vert $.

According to the GU method, we consider the constraint (\ref{pcei}) as the first-class
constraint (fcc) and the remaining constraint (\ref{pgei}) as the
corresponding canonical gauge condition. Further, we redefine the
first-class constraint as
\begin{equation}
G\equiv -\frac{1}{m^{2}}\left( \partial _{i}p^{i}-b\varepsilon
^{0ij}\partial _{i}A_{j}-m^{2}A^{0}\right) \approx 0.  \label{pcein}
\end{equation}%
The other choice, considering the constraint (\ref{pgei}) as the first-class constraint
 and the constraint (\ref{pcei}) as the corresponding
canonical gauge condition, yields a path integral that cannot be written
(after integrating out auxiliary variables) in a manifestly covariant form
\cite{vyt3,IJMPAnoi}. The next step of the GU approach is represented by
the construction of a first-class Hamiltonian with respect to the constraint (\ref{pcein})
\begin{equation}
H_{GU} =H_{c}-\chi ^{(1) }[ G,H_{c}] +\frac{1}{2}%
\chi ^{(1) }\chi ^{( 1) }[ G,[ G,H_{c}] ] -\cdots.
\end{equation}%
The concrete form of first-class Hamiltonian, $H_{GU}$ is given by
\begin{eqnarray}
H_{GU}&=&\int d^{2}x\left[ -\frac{1}{2a}p_{i}p^{i}+\frac{a}{4}\partial
_{\lbrack i}A_{j]}\partial ^{\lbrack i}A^{j]}+\frac{b}{a}\varepsilon
_{0ij}A^{i}p^{j}\right.\nonumber\\
&&-\frac{b^{2}}{2a}A_{i}A^{i}-\frac{m^{2}}{2}A_{0}A^{0}-A_{0}\left( \partial _{i}p^{i}-b\varepsilon ^{0ij}\partial_{i}A_{j}-m^{2}A^{0}\right)\nonumber\\
&&\left.+\frac{1}{2}\left( \frac{1}{m}\partial _{i}p_{0}+mA_{i}\right)\left( \frac{1}{m}\partial ^{i}p_{0}+mA^{i}\right)\right] .\label{phamclsI}
\end{eqnarray}%
It can be verified that the Hamiltonian gauge algebra relation is given by
\begin{equation}
[G,H_{GU}] =0.
\end{equation}
The equations of motion are
\begin{eqnarray}
\dot{A}_{0} &=&-\partial _{i}\left( A^{i}+\frac{1}{m^{2}}\partial
^{i}p^{0}\right) , \label{e1}\\
\dot{A}_{i} &=&-\frac{1}{a}\left(p_{i}+b\varepsilon _{0ij}A^{j}\right)+\partial
_{i}A_{0}+\frac{1}{m^{2}}\partial_{i}\Lambda, \label{e3}\\
\dot{p}^{0} &=&\partial _{i}p^{i}-b\varepsilon ^{0ij}\partial
_{i}A_{j}-m^{2}A^{0}-\Lambda, \label{e2}\\
\dot{p}^{i} &=&a\partial _{j}\partial ^{\lbrack j}A^{i]}-\frac{b}{a}%
\varepsilon _{0ij}\left(p^{j}+b\varepsilon ^{0jk}A_{k}\right)\nonumber\\
&&-m^{2}\left(A^{i}+\frac{1}{m^{2}}\partial ^{i}p^{0}\right)-b\varepsilon
_{0ij}\partial ^{j}A^{0}-\frac{b}{m^2}\varepsilon^0ij\partial_{j}\Lambda,\label{e4}
\end{eqnarray}
where $\Lambda$ is an arbitrary function. Under the canonical gauge condition $p^{0}\approx 0$, ($\Lambda=0$) the equations (\ref{e1})--(\ref{e4}) return to the
equations of motion for the MCS-Proca model. The number of physical degrees of freedom of the GU system is equal to%
\begin{eqnarray}
\mathcal{N}_{GU} &=&( 6\ \rm{canonical\ variables}-2\times 1\ \rm{fcc}) /2  \nonumber \\
&=&2=\mathcal{N}_{O}.  \label{dofgu}
\end{eqnarray}%
The original second-class theory and respectively the gauge-unfixed system
are classically equivalent since they possess the same number of physical
and, moreover, the corresponding algebras of classical observables are
isomorphic. Consequently, the two systems become equivalent at the level of
the path integral quantization, which allows us to replace the Hamiltonian
path integral of the MCS-Proca model with that of the gauge-unfixed
first-class system
\begin{eqnarray}\label{piGU}
Z_{GU}&=&\int \mathcal{D}\left( A_{\mu},p^{\mu},\lambda \right) \mu\left( \left[ A_{\mu}\right]\right)\exp\left\{\mathrm{i}\int d^{3}x\left[ \left( \partial _{0}A_{\mu}\right) p^{\mu}\right.\right.\nonumber\\
&&\left.\left.-\mathcal{H}_{GU}+\frac{1}{m^{2}}\lambda \left( \partial^{i}p_{i}-b\varepsilon^{0ij}\partial _{i}A_{j}-m^{2}A^{0}\right) \right] \right\} ,
\end{eqnarray}
where the integration measure `$\mu\left(\left [ A_{\mu}\right]\right)$' associated with the model subject to the first-class constraint (\ref%
{pcein}) includes some suitable canonical gauge conditions and it is chosen
such that path integral (\ref{piGU}) is convergent \cite{FHP}.

Performing in the path integral the notation
\begin{equation}
\bar{A}_{0}=A_{0}+\frac{1}{m^{2}}\lambda ,  \label{ptransf2}
\end{equation}%
and partial integrations over the momentum $p^{i}$ and field $A_{0}$, the
argument of the exponential takes the form%
\begin{eqnarray}\label{pactot3}
S_{GU}&=&\int d^{3}x\left[ -\frac{a}{4}\partial_{[i}A_{j]}\partial
^{[i}A^{j]}-\frac{a}{2}\left( \partial _{0}A_{i}-\partial _{i}\bar{A}_{0}\left( \partial ^{0}A^{i}-\partial ^{i}\bar{A}^{0}\right)\right)\right.\nonumber\\
&&-b\varepsilon _{0ij}\bar{A}^{0}\partial ^{i}A^{j}-b\varepsilon_{i0j}A^{i}\partial ^{0}A^{j}-b\varepsilon _{ij0}A^{i}\partial ^{j}\bar{A}^{0}\nonumber\\
&&-\frac{1}{2}\left( \frac{1}{m}\partial_{i} p_{0} +mA_{i}\right)\left(\frac{1}{m}\partial ^{i} p^{0}+mA^{i}\right)\nonumber\\
&&\left.-\frac{1}{2}\left( \frac{1}{m}\partial _{0} p_{0}+m\bar{A}_{0}\right)\left( \frac{1}{m}\partial ^{0} p^{0}+m\bar{A}^{0}\right) \right].
\end{eqnarray}
In terms of the notation $\varphi=-\frac{1}{m}p^{0}$, the last functional reads as%
\begin{eqnarray}\label{pacttot3.1}
S_{GU}&=&\int d^{3}x\left[ -\frac{a}{4}\partial _{[\mu }\bar{A}%
_{\nu ]}\partial ^{[\mu }\bar{A}^{\nu ]}-b\varepsilon _{\mu \nu \rho }%
\bar{A}^{\mu }\partial ^{\nu }\bar{A}^{\rho }\right.\nonumber\\
&&\left.-\frac{1}{2}\left( \partial _{\mu }\varphi -m\bar{A}_{\mu }\right)
\left( \partial ^{\mu }\varphi -m\bar{A}^{\mu }\right) \right] ,
\end{eqnarray}%
where  $\bar{A}_{\mu }\equiv\left\{ \bar{A}_{0},A_{i}\right\}$
 and describes a St\"{u}ckelberg coupling between the scalar field $\varphi $ and the $1$-form $%
\bar{A}_{\mu }$ \cite{stueck}. The scalar field $\varphi$ play the role of St\"{u}ckelberg scalar. Using the extended phase space method \cite{FS,bfgen,bfgen1,battyut} in \cite{kor,stuck5,stuck3,stuck2,stuck4} a similar result (for $a=0$ or $b=0$) has been obtained. The extrafield of the extended phase space method was identified with St\"{u}ckelberg scalar. In contrast, in the GU approach we find that to St\"{u}ckelberg scalar corresponds $-\frac{1}{m}p^{0}$, where $p^{0}$ is canonical momentum conjugated with the original field $A_{0}$.

In the following we prove that starting from the Hamiltonian path integral of the gauge
system (\ref{pacttot3.1}) with a suitable gauge we recover the MCS-Proca model. The
canonical analysis of the model described by the Lagrangian action (\ref{actmcs})
displays the first-class constraints
\begin{equation}
G_{1} \equiv p^{0}\approx 0,\quad G_{2} \equiv \partial _{i}p^{i}-b\varepsilon ^{0ij}\partial
_{i}A_{j}-mp\approx 0,  \label{r2}
\end{equation}%
and the Hamiltonian
\begin{eqnarray}
H &=&\int d^{2}x\left[ -\frac{1}{2a}p_{i}p^{i}-A_{0}\partial _{i}p^{i}+\frac{%
a}{4}\partial _{\lbrack i}A_{j]}\partial ^{\lbrack i}A^{j]}+b\varepsilon _{0ij}A^{0}\partial ^{i}A^{j}+\frac{b}{a}\varepsilon
_{0ij}A^{i}p^{j}\right.
\nonumber \\
&&\left. -\frac{b^{2}}{2a}A_{i}A^{i}-\frac{1}{2}p^{2}+mA^{0}p+\frac{1}{2}\left( \partial _{i}\varphi
-mA_{i}\right) \left( \partial ^{i}\varphi -mA^{i}\right) \right] ,
\label{r3}
\end{eqnarray}%
where $\left\{ p^{\mu },p\right\} $ are the canonical momenta conjugated
with the fields $\left\{ A_{\mu },\varphi \right\} $. Taking
\begin{equation}
C^{1} \equiv \varphi \approx 0,  \quad
C^{2} \equiv -p+mA_{0}\approx 0,  \label{r5}
\end{equation}%
as the unitary gauge-fixing conditions, the Hamiltonian path integral is
given by%
\begin{eqnarray}
Z &=&\int \mathcal{D}\left( A_{\mu },p^{\mu },\varphi ,p\right) \delta
\left( G_{1}\right) \delta \left( G_{2}\right) \delta \left( C^{1}\right)
\delta \left( C^{2}\right)   \nonumber \\
&&\times \exp\left\{ \mathrm{i}\int d^{3}x\left[ \left( \partial _{0}A_{\mu
}\right) p^{\mu }+\left( \partial _{0}\varphi \right) p-\mathcal{H}\right] \right\}.
\label{r6}
\end{eqnarray}%
Integrating over the momentum $p^{0}$ and fields  $\{\varphi $, $A_{0}\}$ and representing $\delta
\left( \partial _{i}p^{i}-b\varepsilon ^{0ij}\partial _{i}A_{j}-mp\right) $
in the form of integral functional
\begin{equation}
\int \mathcal{D}\lambda \exp \left\{\mathrm{i}\int d^{3}x\left[\lambda \left( \partial
_{i}p^{i}-b\varepsilon ^{0ij}\partial _{i}A_{j}-mp\right)\right]\right\} ,  \label{r7}
\end{equation}%
the path integral takes the form%
\begin{eqnarray}
Z &=&\int \mathcal{D}\left( A_{i},p^{i},p,\lambda \right) \exp \left\{\mathrm{i}%
\int d^{3}x\left[ \left( \partial _{0}A_{i}\right) p^{i}+\frac{1}{2a}p_{i}p^{i}+\frac{1}{m}p\partial _{i}p^{i}\right.\right.   \nonumber
\\
&& -\frac{a}{4}\partial
_{\lbrack i}A_{j]}\partial ^{\lbrack i}A^{j]}-\frac{b}{m}\varepsilon _{0ij}p\partial ^{i}A^{j}-\frac{b}{a}\varepsilon
_{0ij}A^{i}p^{j}+\frac{b^{2}}{2a}A_{i}A^{i} \nonumber \\
&&\left.\left.-\frac{1}{2}p^{2} -\frac{m^{2}}{2}A_{i}A^{i}+\lambda \left( \partial
_{i}p^{i}-b\varepsilon ^{0ij}\partial _{i}A_{j}-mp\right) \right]\right\} .
\label{r8}
\end{eqnarray}%
Performing in the path integral the notation
\begin{equation}
A_{0}=\frac{1}{m}p+\lambda ,  \label{r9}
\end{equation}%
the argument of the exponential becomes%
\begin{eqnarray}
Z &=&\int \mathcal{D}\left( A_{\mu },p^{i},p\right) \exp\left\{ \mathrm{i}\int
d^{3}x\left[ \left( \partial _{0}A_{i}\right) p^{i}+\frac{1}{2a}p_{i}p^{i}\right.\right.   \nonumber \\
&&-\frac{a}{4}\partial _{\lbrack i}A_{j]}\partial
^{\lbrack i}A^{j]}-\frac{b}{a}\varepsilon _{0ij}A^{i}p^{j} -\frac{b^{2}}{2a}A_{i}A^{i}+\frac{1}{2}p^{2} \nonumber \\
&&\left.\left.-\frac{m^{2}}{2}A_{i}A^{i} +A_{0}\left( \partial _{i}p^{i}-b\varepsilon ^{0ij}\partial
_{i}A_{j}-mp\right) \right]\right\} .  \label{r10}
\end{eqnarray}%
After integration over the momenta $p^{i}$ and $p$, we find that the argument of the exponential is just the MCS-Proca Lagrangian
\begin{eqnarray}
Z&=&\int \mathcal{D}A_{\mu }\exp \left\{\mathrm{i}\int d^{3}x\left( -\frac{a}{4}%
\partial _{\lbrack \mu }A_{\nu ]}\partial ^{\lbrack \mu }A^{\nu
]}\right.\right.\nonumber\\
&&\left.\left.-b\varepsilon _{\mu \nu \rho }A^{\mu }\partial ^{\nu }A^{\rho }-\frac{m^{2}%
}{2}A_{\mu }A^{\mu }\right)\right\} .  \label{r11}
\end{eqnarray}

The MCS-Proca model can be correlated to another first-class theory whose
field spectrum comprise two types of $1$-form gauge fields. For this purpose
we consider the following fields/momenta combinations%
\begin{equation}
\mathcal{P}_{i} \equiv p_{i}+b\varepsilon _{0ij}A^{j},\quad  \mathcal{F}_{i} \equiv A_{i}+\frac{1}{m^{2}}\partial _{i}p_{0},\quad
\mathcal{F}_{0} \equiv A_{0},  \label{x2.3}
\end{equation}%
which are in (strong) involution with the first-class constraint (\ref{pcein}%
)
\begin{equation}
\left[ \mathcal{P}_{i},G\right] =\left[ \mathcal{F}_{i},G\right] =\left[
\mathcal{F}_{0},G\right] =0.  \label{x3}
\end{equation}%
We observe that the first-class Hamiltonian (\ref{phamclsI}) can be written
in terms of these gauge invariant quantities as%
\begin{eqnarray}
H_{GU}&=&\int d^{2}x\left[ -\frac{1}{2a}\mathcal{P}_{i}\mathcal{P}^{i}+%
\frac{a}{4}\partial _{[i}\mathcal{F}_{j]}\partial ^{[ i}%
\mathcal{F}^{j]}\right.\nonumber\\
&&\left.+\frac{m^{2}}{2}\mathcal{F}_{i}\mathcal{F}^{i}-\frac{m^{2}}{2}%
\mathcal{F}_{0}\mathcal{F}^{0}+m^{2}\mathcal{F}_{0}G\right] .  \label{x1}
\end{eqnarray}%
By direct computation we find that $\mathcal{F}_{\mu }\equiv \left\{
\mathcal{F}_{0},\mathcal{F}_{i}\right\} $  satisfy the equations%
\begin{eqnarray}
\partial ^{\nu }\partial _{\lbrack \nu }\mathcal{F}_{0]}&=&\frac{m^{2}}{a}\mathcal{F}_{0}+\frac{2b}{a}\varepsilon _{0ij}\partial^{i}\mathcal{F}^{j}+\mathcal{O}\left( G\right) ,  \label{x4.1} \\
\partial ^{\nu }\partial _{\lbrack \nu }\mathcal{F}_{i]}&=&\frac{m^{2}}{a}%
\mathcal{F}_{i}+\frac{2b}{a^{2}}\varepsilon _{0ij}\mathcal{P}^{j}+\mathcal{O%
}\left( G\right),  \label{x4.2}
\end{eqnarray}
and is divergenceless%
\begin{equation}
\partial ^{\mu }\mathcal{F}_{\mu }=0. \label{x6}
\end{equation}
Enlarging the phase space by adding some bosonic canonical variables $\left\{V^{\mu
},P_{\mu }\right\}$, we can write the solution to the Eq. (\ref{x6}) as
\begin{equation}
\mathcal{F}_{\mu }=-\frac{1}{m}\varepsilon _{\mu \nu \rho }\partial ^{\nu
}V^{\rho }.  \label{x7}
\end{equation}%
When we replace the solution (\ref{x7}) in the first-class constraint (\ref{pcein}), the constraint takes the form
\begin{equation}
-\frac{1}{m^{2}}\left( \partial ^{i}p_{i}-b\varepsilon _{0ij}\partial
^{i}A^{j}+m\varepsilon _{0ij}\partial ^{i}V^{j}\right) \approx 0,  \label{x9}
\end{equation}%
and remains first-class. From the gauge transformation of the quantity $\partial _{i}p_{0}$, we obtain that%
\begin{equation}
\partial _{i}p_{0}=m\varepsilon _{0ij}P^{j}.  \label{x11}
\end{equation}%
Using the relations (\ref{x7}) and (\ref{x11}) in the first-class Hamiltonian (\ref{phamclsI}), we obtain for the
first-class Hamiltonian the following form
\begin{eqnarray}\label{x12}
H_{GU}^{\prime }&=&\int d^{2}x\left[\frac{a}{4}\partial _{[ i}A_{j]}\partial^{[ i}A^{j]}-\frac{1}{4}\partial _{[ i}V_{j]}\partial ^{[ i}V^{j]} \right.\nonumber\\
&&-\frac{1}{2a}\left(p_{i}+b\varepsilon _{0ij}A^{j}\right) \left( p^{i}+b\varepsilon^{0ik}A_{k}\right)\nonumber\\
&&+\frac{m^{2}}{2}\left( A_{i}+\frac{1}{m}\varepsilon _{0ij}P^{j}\right)\left( A_{i}+\frac{1}{m}\varepsilon ^{0ik}P_{k}\right)\nonumber\\
&&\left.+\frac{1}{2m}\varepsilon _{0ij}\partial ^{[ i}V^{j]}\left(\partial _{k}p^{k}-b\varepsilon _{0kl}\partial ^{k}A^{l}+m\varepsilon _{0kl}\partial ^{k}V^{l}\right) \right] .
\end{eqnarray}
In this moment we have a dynamical system with the phase space locally
parameterized by $\left\{A_{i},p^{i},V^{\mu },P_{\mu }\right\}$, subject to the
first-class constraint (\ref{x9}) and too many degrees of freedom%
\begin{eqnarray}
\mathcal{N}_{GU}^{\prime } &=&( 10\ \rm{canonical\ variables}-2\times
1\ \rm{fcc}) /2  \nonumber \\
&=&4\neq \mathcal{N}_{GU}.
\end{eqnarray}%
In order to cut the two extra degrees of freedom, we impose in addition to the first-class constraint (\ref{x9}) two supplementary first-class constraints
\begin{equation}
-\partial _{i}P^{i}\approx 0,\qquad P^{0}\approx 0,  \label{x10}
\end{equation}%
and we obtain a first-class system with a right number of physical degrees
of freedom%
\begin{eqnarray}
\mathcal{N}_{GU}^{\prime } &=&( 10\ \rm{canonical\ variables}-2\times
3\ \rm{fcc}) /2  \nonumber \\
&=&2=\mathcal{N}_{GU}.  \label{dofgu'}
\end{eqnarray}%
Since the number of physical degrees of freedom is the same for both first-class theories and for each of them we are able to identify a set of fundamental classical observables such that they are in one-to-one
correspondence and possess the same Poisson brackets, the first-class
theories are equivalent. As a result, the GU and the first-class systems
remain equivalent also at the level of the Hamiltonian path integral
quantization. This further implies that the first-class system is completely
equivalent with the original second-class theory. Due to this equivalence we
can replace the Hamiltonian path integral of MCS-Proca model with that one
associated with the first-class system%
\begin{eqnarray}
Z^{\prime }&=&\int \mathcal{D}\left( A_{i},V^{\mu },p^{i},P_{\mu },\lambda's\right) \mu \left( [A_{i}],[V^{\mu}]\right)\nonumber\\
&&\times\exp\left\{\mathrm{i}\int d^{3}x\left[ \left( \partial _{0}A_{i}\right)
p^{i}+\left( \partial _{0}V^{\mu }\right) P_{\mu }-\mathcal{H}_{GU}^{\prime}\right.\right.\nonumber\\
&&\left.\left.+\lambda ^{(1)}\partial _{i}P^{i}-\lambda ^{(2)}P^{0}+\frac{1}{m^{2}}\lambda
\left( \partial _{i}p^{i}-b\varepsilon _{0ij}\partial ^{i}A^{j}+m\varepsilon _{0ij}\partial ^{i}V^{j}\right)\right]\right\} .
\label{x16}
\end{eqnarray}%
If we perform in path integral the partial integrations over $\left\{V^{0},p_{i},P_{\mu},\lambda ^{(2)}\right\}$ and use the notations
\begin{equation}
\bar{A}_{0}=\frac{1}{m^{2}}\left( \lambda -\frac{%
m}{2}\varepsilon _{0ij}\partial ^{[ i}V^{j]}\right), \qquad \bar{V}_{0}=\lambda^{(1)},  \label{x14}
\end{equation}%
the argument of the exponential becomes%
\begin{eqnarray}
S_{GU}^{\prime }&=& \int d^{3}x\left[ -\frac{a}{4}\partial _{[i}A_{j]}\partial ^{[ i}A^{j]}-\frac{a}{2}\left(\partial_{0}A_{i}-\partial _{i}\bar{A}_{0}\right)\left(\partial^{0}A^{i}-\partial ^{i}\bar{A}^{0}\right)
\right.\nonumber\\
&&-b\varepsilon _{0ij}\bar{A}^{0}\partial ^{i}A^{j}-b\varepsilon_{i0j}A^{i}\partial ^{0}A^{j}-b\varepsilon_{ij0}A^{i}\partial ^{j}\bar{A}^{0}\nonumber\\
&&+\frac{1}{4}\partial _{[ i}V_{j]}\partial ^{[i}V^{j]}+\frac{1}{2}\left( \partial _{0}V_{i}-\partial _{i}\bar{V}_{0}\right) \left( \partial ^{0}V^{i}-\partial ^{i}\bar{V}^{0}\right)\nonumber \\
&&\left.+m\varepsilon _{0ij}\bar{A}^{0}\partial ^{i}V^{j}+m\varepsilon
_{i0j}A^{i}\partial ^{0}V^{j}+m\varepsilon
_{ij0}A^{i}\partial ^{j}\bar{V}^{0}\right] .\label{x17}
\end{eqnarray}
The argument of the exponential takes a manifestly Lorentz-covariant form%
\begin{eqnarray}
S_{GU}^{\prime }&=&\int d^{3}x\left( -\frac{a}{4}\partial _{[ \mu }%
\bar{A}_{\nu ]}\partial ^{[ \mu }\bar{A}^{\nu ]}-b\varepsilon _{\mu
\nu \rho }\bar{A}^{\mu }\partial ^{\nu }\bar{A}^{\rho }\right.\nonumber\\
&&\left.+\frac{1}{4}\partial _{[ \mu }\bar{V}_{\nu ]}\partial
^{[ \mu }\bar{V}^{\nu ]}+m\varepsilon _{\mu \nu \rho }\bar{A}^{\mu
}\partial ^{\nu }\bar{V}^{\rho }\right), \label{x19}
\end{eqnarray}%
where $\bar{A}_{\mu }\equiv \left\{ \bar{A}_{0},A_{i}\right\} $ and $\bar{V}%
_{\mu }\equiv \left\{ \bar{V}_{0},V_{i}\right\} $. The functional (\ref{x19}%
) associated with the first-class system describes a CS coupling
between the two $1$-forms, $\bar{A}_{\mu }$ and $\bar{V}_{\mu }$ \cite{BCS}.

\section{The higher derivative MCS-Proca model\label{sec3}}

\subsection{Hamiltonian analysis of the MECS-Proca model\label{sec3.1}}

 The starting point of the approach developed in \cite{BMP} consists in converting the original higher
derivative theory to an equivalent first order theory by introducing new
fields to account for higher derivative terms. To pass from the higher derivative theory to a first order one, we define the variables $B_{\mu }$ as%
\begin{equation}
B_{\mu }=\partial _{0}A_{\mu },  \label{y1.1}
\end{equation}%
and enforce the Lagrangian constraints%
\begin{equation}
B_{\mu }-\partial _{0}A_{\mu }=0,
\end{equation}%
by Lagrange multiplier $\xi ^{\mu }$%
\begin{eqnarray}
\mathcal{L}&=&-\frac{a}{4}\partial _{[ i}A_{j]}\partial ^{[i}A^{j]}-\frac{a}{2}\left( B_{i}-\partial _{i}A_{0}\right) \left(
B^{i}-\partial ^{i}A^{0}\right)\nonumber\\
&&+\frac{1}{2b}\varepsilon _{0ij}\left( \partial _{0}B^{0}+\partial
_{k}\partial ^{k}A^{0}\right) \partial ^{i}A^{j}+\frac{1}{2b}\varepsilon
_{i0j}\left( \partial _{0}B^{i}+\partial _{k}\partial ^{k}A^{i}\right)B^{j}\nonumber\\
&&+\frac{1}{2b}\varepsilon
_{ij0}\left( \partial _{0}B^{i}+\partial _{k}\partial ^{k}A^{i}\right) \partial ^{j}A^{0}-\frac{m^{2}}{2}A_{\mu }A^{\mu
}+\xi ^{\mu }\left( B_{\mu }-\partial _{0}A_{\mu }\right) .  \label{y1.2}
\end{eqnarray}
From the definitions of the canonical momenta $\left\{ \Pi _{\mu },p^{\mu
},\pi ^{\mu }\right\} $ conjugate to the fields $\left\{ \xi ^{\mu },A_{\mu
},B_{\mu }\right\} $
\begin{equation}
\Pi _{\mu } =\frac{\partial L}{\partial \dot{\xi}^{\mu }},\quad  p^{\mu } =\frac{\partial L}{\partial \dot{A}_{\mu }}, \quad
\pi ^{\mu } =\frac{\partial L}{\partial \dot{B}_{\mu }},  \label{a3}
\end{equation}%
we obtain the primary constraints%
\begin{eqnarray}
&&\Phi _{\mu }^{(\xi )}\equiv \Pi _{\mu }\approx 0,  \label{y2.1} \\
&&\Phi ^{(A)\mu }\equiv p^{\mu }+\xi ^{\mu }\approx 0,  \label{y2.2} \\
&&\Phi _{i}^{(B)}\equiv \pi _{i}+\frac{1}{2b}\varepsilon _{0ij}\left(
B^{j}-\partial ^{j}A^{0}\right) \approx 0,  \label{a6} \\
&&\Phi ^{(B)}\equiv \pi _{0}-\frac{1}{2b}\varepsilon _{0ij}\partial
^{i}A^{j}\approx 0.  \label{a7}
\end{eqnarray}%
If we write the primary constraints (\ref{a6})--(\ref{a7}) in an equivalent form%
\begin{eqnarray}
&&\Phi _{i}^{\prime (B)}\equiv \pi _{i}+\frac{1}{2b}\varepsilon _{0ij}\left(
B^{j}-\partial ^{j}A^{0}\right) -\frac{1}{2b}\varepsilon _{0ij}\partial
^{j}\Pi ^{0}\approx 0,  \label{y2.4} \\
&&\Phi ^{\prime (B)}\equiv \pi _{0}-\frac{1}{2b}\varepsilon _{0ij}\partial
^{i}A^{j}-\frac{1}{2b}\varepsilon _{0ij}\partial ^{i}\Pi ^{j}\approx 0,
\label{y2.3}
\end{eqnarray}%
the nonvanishing elements of the algebra of the primary constraints (pc) are%
\begin{eqnarray}
&&\left[ \Phi _{\mu }^{(\xi )}(x),\Phi ^{(A)\nu }(y)\right]
_{x_{0}=y_{0}}=-\delta _{\mu }^{\nu }\delta ^{2}(\mathbf{x}-\mathbf{y}),
\label{a13} \\
&&\left[ \Phi _{i}^{\prime (B)}(x),\Phi _{j}^{\prime (B)}(y)\right]
_{x_{0}=y_{0}}=\frac{1}{b}\varepsilon _{0ij}\delta ^{2}(\mathbf{x}-\mathbf{y}%
).  \label{a14}
\end{eqnarray}%
The canonical Hamiltonian is given by
\begin{eqnarray}
H_{c} &=&\int d^{2}x\left. \left( \Pi _{\mu }\dot{\xi}^{\mu }+p^{\mu }\dot{A}%
_{\mu }+\pi ^{\mu }\dot{B}_{\mu }-\mathcal{L}\right) \right\vert _{\left\{
pc
\right\} }  \nonumber \\
&=&\int d^{2}x\left[ \frac{a}{4}\partial _{\lbrack i}A_{j]}\partial ^{\lbrack i}A^{j]}+%
\frac{a}{2}\left( B_{i}-\partial _{i}A_{0}\right) \left( B^{i}-\partial
^{i}A^{0}\right) \right.   \nonumber \\
&&-\frac{1}{2b}\varepsilon _{0ij}\left( \partial _{k}\partial
^{k}A^{0}\right) \partial ^{i}A^{j}-\frac{1}{2b}\varepsilon _{i0j}\left(
\partial _{k}\partial ^{k}A^{i}\right) B^{j}  \nonumber \\
&&\left. -\frac{1}{2b}\varepsilon _{ij0}\left( \partial _{k}\partial
^{k}A^{i}\right) \partial ^{j}A^{0}-\xi ^{\mu }B_{\mu }+\frac{m^{2}}{2}%
A_{\mu }A^{\mu }\right] ,  \label{y3.1}
\end{eqnarray}%
and total Hamiltonian is%
\begin{equation}
H_{T}=H_{c}+\int d^{2}x\left(u^{\left( \xi \right) \mu }\Phi _{\mu }^{(\xi )}+u_{\mu
}^{\left( A\right) }\Phi ^{(A)\mu }+u^{\left( B\right) i}\Phi _{i}^{\prime
(B)}+u^{\left( B\right) }\Phi ^{\prime (B)}\right),  \label{a15}
\end{equation}%
where $\left\{ u^{\left( \xi \right) \mu },u_{\mu }^{\left( A\right)
},u^{\left( B\right) i},u^{\left( B\right) }\right\} $ are Lagrange
multipliers.

The consistency of the primary constraints (\ref{y2.1}), (\ref{y2.2}), (\ref{y2.4}) leads to the
determination of the Lagrange multipliers $\left\{ u^{\left( \xi \right) \mu },u_{\mu
}^{\left( A\right) },u^{\left( B\right) i}\right\} $, while the consistency of the
remaining primary constraint $\Phi ^{\prime (B)}\approx 0$ generate the
secondary constraint%
\begin{equation}
\Phi _{II}^{(B)}\equiv \xi _{0}-\frac{1}{2b}\varepsilon _{0ij}\partial
^{i}B^{j}\approx 0.  \label{y4}
\end{equation}%
The consistency of the secondary constraint yields the tertiary
constraint%
\begin{equation}
\Phi _{III}^{(B)}\equiv \partial _{i}\xi ^{i}+m^{2}A_{0}-\frac{1}{2b}%
\varepsilon _{0ij}\partial _{k}\partial ^{k}\partial ^{i}A^{j}\approx 0.
\label{y5}
\end{equation}%
Conserving the constraint $\Phi _{III}^{(B)}\approx 0$ we get the quartic
constraint
\begin{equation}
\Phi _{IV}^{(B)}\equiv m^{2}\partial _{i}A^{i}+m^{2}B_{0}\approx 0.
\label{y6}
\end{equation}%
The consistency condition of the quartic constraint $\Phi _{IV}^{(B)}\approx
0$ determines the multiplier $u^{\left( B\right) }$ and no more new
constraint is produced.

The constraints (\ref{y2.1}), (\ref{y2.2}), (\ref{y2.4}), (\ref{y2.3}) and (\ref{y4})--(\ref{y6}) are second-class
and irreducible. The nonzero Poisson brackets among the constraints functions
read as%
\begin{eqnarray}
&&\left[ \Phi _{\mu }^{( \xi ) }( x) ,\Phi ^{(
A) \nu }( y) \right] _{x_{0}=y_{0}}=-\delta _{\mu }^{\nu
}\delta ^{2}( \mathbf{x}-\mathbf{y}) , \\
&&\left[ \Phi _{\mu }^{( \xi ) }( x) ,\Phi _{II}^{(
B) }( y) \right] _{x_{0}=y_{0}}=-\delta _{\mu
}^{0}\delta ^{2}( \mathbf{x}-\mathbf{y}) , \\
&&\left[ \Phi _{\mu }^{( \xi ) }( x) ,\Phi
_{III}^{( B) }( y) \right] _{x_{0}=y_{0}} =\delta
_{\mu }^{i}\partial _{i}\delta ^{2}( \mathbf{x}-\mathbf{y}) , \\
&&\left[ \Phi ^{( A) 0 }( x) ,\Phi _{III}^{(
B) }( y) \right] _{x_{0}=y_{0}} =-m^{2}\delta ^{2}( \mathbf{x}-\mathbf{y}) , \\
&&\left[ \Phi ^{( A) i }( x) ,\Phi _{III}^{(
B) }(y) \right] _{x_{0}=y_{0}} =\frac{1}{2b}\varepsilon ^{0ij}\partial _{k}\partial
^{k}\partial _{j} \delta ^{2}( \mathbf{x}-\mathbf{y}) , \\
&&\left[ \Phi ^{(A) \mu }(x) ,\Phi _{IV}^{(B) }(y) \right] _{x_{0}=y_{0}} =m^{2}\delta _{i}^{\mu
}\partial ^{i}\delta ^{2}( \mathbf{x}-\mathbf{y}) , \\
&&\left[ \Phi ^{\prime(B) }(x) ,\Phi _{IV}^{(B)
}(y) \right] _{x_{0}=y_{0}} =-m^{2}\delta ^{2}(\mathbf{x}-%
\mathbf{y}) , \\
&&\left[ \Phi _{i}^{\prime(B) }(x) ,\Phi _{j}^{\prime(B) }(y) \right] _{x_{0}=y_{0}} =\frac{1}{b}\varepsilon _{0ij}\delta ^{2}( \mathbf{x}-\mathbf{y}) , \\
&&\left[ \Phi _{i}^{\prime(B) }(x) ,\Phi _{II}^{(B) }(y) \right] _{x_{0}=y_{0}} =\frac{1}{b}\varepsilon
_{0ij}\partial ^{j}\delta ^{2}( \mathbf{x}-\mathbf{y}) .
\end{eqnarray}
The number of physical degrees of freedom of the original system is equal to%
\begin{eqnarray}
\bar{\mathcal{N}}_{O} &=( 18\ \rm{canonical\ variables}-12\ \rm{scc%
}) /2  \nonumber \\
&=3.  \label{dofoemcs}
\end{eqnarray}%
We notice that the number of physical degrees of freedom of the extended
model is higher than the number of physical degrees of freedom of the
MCS-Proca model
\begin{equation}
\bar{\mathcal{N}}_{O}>\mathcal{N}_{O},
\end{equation}
This result was expected due to the higher derivative nature of the
MECS-Proca model. In addition the number of physical degrees of freedom of MECS-Proca model coincides with that
of the ECSPMG theory.

The analyze of the excitations and mass counts of the MECS-Proca model
reveal the fact that if the sign of the Maxwell term is the usual one then
the excitation masses will be complex, with the wrong sign the reality of the
excitation masses will be restored for a known condition satisfied by parameters $%
b$ and $m$, but the model faces ghost problems. The action (\ref{y1}) can be
rewritten in terms of the the transverse operator
 $ \theta _{\mu \nu }=\sigma_{\mu \nu }-\frac{\partial _{\mu }\partial _{\nu }}{\square} $,
  longitudinal operators
   $ \omega _{\mu \nu }=\frac{\partial _{\mu }\partial _{\nu }}{\square} $
   and the operator associated with the topological term $S_{\mu \nu
}=\varepsilon _{\mu \rho \nu }\partial ^{\rho }$ like
\begin{equation}
S=\int d^{3}x\frac{1}{2}A^{\mu }\mathcal{O}_{\mu\nu}A^{\nu },  \label{p1}
\end{equation}%
where $\mathcal{O}_{\mu\nu}= \left( a \square-m^{2}\right) \theta
_{\mu \nu }-m^{2}\omega _{\mu \nu }+\frac{1}{b}\square S_{\mu \nu }$. The propagator in the momentum space for the MECS-Proca model is
\begin{equation}
\mathcal{P}_{\mu \nu }=-\frac{ak^{2}+m^{2}}{\left( ak^{2}+m^{2}\right) ^{2}-%
\frac{1}{b^{2}}k^{6}}\theta _{\mu \nu }-\frac{1}{m^{2}}\omega _{\mu \nu }+%
\frac{\frac{1}{b}k^{2}}{\left( ak^{2}+m^{2}\right) ^{2}-\frac{1}{b^{2}}k^{6}}%
S_{\mu \nu }.  \label{p2}
\end{equation}%
Taking into consideration that only $\theta $-component of the propagator
\begin{equation}
\mathcal{P}^{\left( \theta \right) }=-\frac{ak^{2}+m^{2}}{\left(
ak^{2}+m^{2}\right) ^{2}-\frac{1}{b^{2}}k^{6}},
\end{equation}%
contributes to the current-current transition amplitude, we study the
residues at each simple pole of the $\mathcal{P}^{\left( \theta \right) }$%
 \cite{AD1,AD2}.

We analyze the roots of the cubic equation%
\begin{equation}
-\frac{1}{b^{2}}\left( k^{2}\right) ^{3}+a^{2}\left( k^{2}\right)
^{2}+2am^{2}k^{2}+m^{4}=0,  \label{p3}
\end{equation}%
whose discriminant is%
\begin{equation}
\mathrm{D}=4\frac{m^{8}}{b^{4}}\left( -a^{3}\frac{b^{2}}{m^{2}}-\frac{27}{4}%
\right) .  \label{p4}
\end{equation}%
For $a=1$ (Maxwell's term with usual sign) the discriminant is less than
zero and the equation has one real root and two complex conjugate roots.
Also, for $a=-1$ (Maxwell's term with the wron sign) the roots of the
equation (\ref{p3}) are complex unless $\frac{b^{2}}{m^{2}}\geq \frac{27}{4}$%
. In the limit case $\frac{b^{2}}{m^{2}}=\frac{27}{4}$ the roots coalesce
and are%
\begin{equation}
k_{1}^{2}=k_{2}^{2}=4k_{3}^{2}=3m^{2}.  \label{p5}
\end{equation}%
Therefore, if $a=-1$ and $\frac{b^{2}}{m^{2}}>\frac{27}{4}$ the equation has
three distinct real roots. In  \cite{PPT} (see also  \cite{DT}) the equation (\ref{p3})
for $a=-1$ was obtained from the pole propagator of the ECSPMG model, where it was noted that if $\frac{b^{2}}{%
m^{2}}>\frac{27}{4}$ then the three distinct real roots are all positive.
The absence of the tachyons in a theory is provided by the existence of only
positive poles, and consequently the MECS-Proca model is free of the
tachyons for $a=-1$ and $\frac{b^{2}}{m^{2}}>\frac{27}{4}$. After the
analyze of the signs of the residues at each simple pole of $\theta $%
-component of the propagator, we obtain that not all residues have the same
sign. The signs of the residues at each simple pole of $\theta $-component
of the propagator tell us whether the ghosts excitations arise and therefore
the MECS-Proca model is plagued by ghosts. We notice that the same problems from the ECSPMG theory
about the presence of ghosts and tachyon excitations are also
present here, in the MECS-Proca model.

\subsection{The construction of the first-class system}

Imposing the constraints (\ref{y2.1})--(\ref{y2.2}) strongly zero and eliminating the
unphysical sector $\{\xi ^{\mu },\Pi _{\mu }\}$, the reduced phase space
being locally parameterized by $\left\{ A_{\mu },B_{\mu },p^{\mu },\pi ^{\mu
}\right\} $, we arrive at a system subject to the second-class constraints%
\begin{eqnarray}
\chi _{i}^{( 1) } &\equiv& \pi _{i}+\frac{1}{2b}\varepsilon
_{0ij}\left(B^{j}-\partial ^{j}A^{0}\right)\approx 0,
\label{y8.1} \\
\chi ^{( 1) } &\equiv& \pi _{0}-\frac{1}{2b}\varepsilon
_{0ij}\partial ^{i}A^{j}\approx 0,  \label{y8.2} \\
\chi ^{( 2) } &\equiv& -p_{0}-\frac{1}{2b}\varepsilon
_{0ij}\partial ^{i}B^{j}\approx 0,  \label{y8.3} \\
\chi ^{( 3) } &\equiv& -\partial _{i}p^{i}+m^{2}A_{0}-\frac{1}{2b}%
\varepsilon _{0ij}\partial _{k}\partial ^{k}\partial ^{i}A^{j}
\approx 0,  \label{y8.4} \\
\chi ^{( 4) } &\equiv& m^{2}\partial _{i}A^{i}+m^{2}B_{0}\approx 0,
\label{y8.5}
\end{eqnarray}
while the canonical Hamiltonian (\ref{y3.1}) takes the form%
\begin{eqnarray}
H_{c}&=&\int d^{2}x\left[ \frac{a}{4}\partial _{ [i}A_{j]}\partial
^{ [i}A^{j]}+\frac{a}{2}\left( B_{i}-\partial _{i}A_{0}\right)\left(
B^{i}-\partial ^{i}A^{0}\right)\right.\nonumber\\
&&-\frac{1}{2b}\varepsilon _{0ij}\left( \partial _{k}\partial
^{k}A^{0}\right) \partial ^{i}A^{j}-\frac{1}{2b}\varepsilon _{i0j}\left(
\partial _{k}\partial ^{k}A^{i}\right) B^{j}\nonumber\\
&&\left.-\frac{1}{2b}\varepsilon _{ij0}\left(
\partial _{k}\partial ^{k}A^{i}\right)  \partial ^{j}A^{0} +p^{\mu }B_{\mu }+\frac{m^{2}}{2}A_{\mu
}A^{\mu }\right] .  \label{y9}
\end{eqnarray}%
The nontrivial Poisson brackets between the constraints functions are listed
bellow%
\begin{eqnarray}
&&\left[ \chi _{i}^{( 1) }( x) ,\chi _{j}^{(1) }( y) \right] _{x_{0}=y_{0}} = \frac{1}{b}\varepsilon _{0ij}\delta ^{2}( \mathbf{x}-\mathbf{y}) ,  \label{y10.1} \\
&&\left[ \chi _{i}^{( 1) }( x) ,\chi ^{( 2)
}( y) \right] _{x_{0}=y_{0}} =\frac{1}{b}\varepsilon _{0ij}\partial
^{j}\delta ^{2}( \mathbf{x}-\mathbf{y}) , \\
&&\left[ \chi ^{( 1) }( x) ,\chi ^{( 4)
}( y) \right] _{x_{0}=y_{0}} =-m^{2}\delta ^{2}( \mathbf{x}-\mathbf{y}) ,  \label{y10.2} \\
&&\left[ \chi ^{( 2) }( x) ,\chi ^{( 3)
}( y) \right] _{x_{0}=y_{0}} =m^{2}\delta ^{2}( \mathbf{x}-\mathbf{y}) , \\
&&\left[ \chi ^{( 3) }( x) ,\chi ^{( 4)
}( y) \right] _{x_{0}=y_{0}} =-m^{2}\partial _{k}\partial ^{k}\delta ^{2}( \mathbf{x}-\mathbf{y}).
\end{eqnarray}%
If we make the following linear combination of the constraints $\chi
^{( 2) }\approx 0$ and $\chi _{i}^{( 1) }\approx 0$
\begin{equation}
\bar{\chi} ^{( 2) }=\chi ^{( 2) }+\partial
^{i}\chi _{i}^{( 1) }\approx
0,  \label{y11}
\end{equation}%
the matrix of the Poisson bracket among the constraints functions becomes%
\begin{equation}
C_{\alpha _{0}\beta _{0}}=\left(
\begin{array}{ccccc}
\frac{1}{b}\varepsilon _{0ij} & \mathbf{0} & \mathbf{0} & \mathbf{0} & \mathbf{0} \\
\mathbf{0} & 0 & 0 & 0 & -1 \\
\mathbf{0} & 0 & 0 & m^{2} & 0 \\
\mathbf{0} & 0 & -m^{2} & 0 & -m^{2}\partial _{k}\partial ^{k} \\
\mathbf{0} & 1 & 0 & m^{2}\partial _{k}\partial ^{k} & 0%
\end{array}\right) .  \label{y12}
\end{equation}%
We notice that the constraints $\chi _{i}^{(1)}\approx 0$  generate a submatrix (of the matrix of the Poisson
brackets among the constraints functions) of maximum rank, therefore they
form an independent subset of second-class constraints. Thus in the sequel we
examine from the point of view of the GU method only the constraints $\chi
_{A}\equiv \left\{ \chi ^{( 1) },\bar{\chi} ^{( 2) },\chi
^{( 3) },\chi ^{( 4) }\right\} \approx 0$.

The second-class constraints set $\chi _{A}\approx 0$ cannot be
straightforwardly separated in two subsets such that one of them being
first-class and the other providing some canonical gauge conditions for the
first-class subset. To make this possible, we write the constraints set in an
equivalent form%
\begin{equation}
\chi _{A}^{\prime }=E_{AB}\chi _{B},  \label{y13}
\end{equation}%
where $E_{AB}$ is an invertible matrix
\begin{equation}
E_{AB}=
\left(
\begin{array}{cccc}
\frac{\partial _{k}\partial ^{k}}{m^{2}} & 0 & -\frac{1}{m^{2}} & 0 \\
0 & 1 & 0 & 0 \\
-1 & 0 & 0 & 0 \\
0 & 0 & 0 & \frac{1}{m^{2}}%
\end{array}\right) .  \label{y14}
\end{equation}%
The concrete form of the constraints $\chi _{A}^{\prime }\approx 0$ is%
\begin{eqnarray}
\chi ^{\prime ( 1) } &\equiv&\frac{1}{m^{2}}\left(\partial _{i}p^{i}-m^{2}A_{0}+\partial
_{k}\partial ^{k}\pi _{0}\right)\approx 0,  \label{y15.1} \\
\chi ^{\prime ( 2) } &\equiv& -p_{0}+\partial _{i}\pi ^{i}\approx
0,  \label{y15.2} \\
\chi ^{\prime ( 3) } &\equiv& -\pi _{0}+\frac{1}{2b}\varepsilon
_{0ij}\partial ^{i}A^{j}\approx 0,  \label{y15.3} \\
\chi ^{\prime ( 4) } &\equiv& \partial _{i}A^{i}+B_{0}\approx 0,
\label{y15.4}
\end{eqnarray}%
with the matrix of the Poisson brackets among the constraints functions
expressed by
\begin{equation}
C_{AB}=\left(
\begin{array}{cccc}
0 & 1 & 0 & 0 \\
-1 & 0 & 0 & 0 \\
0 & 0 & 0 & 1 \\
0 & 0 & -1 & 0%
\end{array}\right).  \label{y16}
\end{equation}%
Examining the structure of the constraints set (\ref{y15.1})--(\ref{y15.4}) we notice that, in the constraints  $\chi ^{\prime( 1) }\approx 0$ and $\chi ^{\prime( 2) }\approx 0$ we find the reminiscence of the structure of the constraints set of MCS-Proca model (\ref{pgei})--(\ref{pcei}), while the constrains $\chi ^{\prime( 3) }\approx 0$ and $\chi ^{\prime( 4) }\approx 0$ have no counterparts. It has proved in Ref. \cite{mitra} that for a dynamical system
subject to the second-class constraints $\left\{ \chi _{\alpha _{0}}\approx 0\right\} _{\alpha _{0}=\overline{1,2M_{0}}}$, the subsets $\left\{ \chi
_{1},\chi _{2},\ldots ,\chi _{M_{0}}\right\} $ and $\left\{ \chi _{1},\chi _{2},\ldots ,\chi _{M_{0}-1},\chi _{M_{0}+1}\right\} $ of the full set of
constraints are first-class sets on $\Sigma _{2M_{0}}$. According to the above, we consider the subset $G_{a}\equiv \left\{ \chi ^{\prime( 1) },\ \chi ^{\prime ( 3) }\right\}$ as the first-class constraints set and the remaining constraints $C_{a}\equiv \left\{ \chi ^{\prime ( 2) },\ \chi ^{\prime
( 4) }\right\} $ as the corresponding canonical gauge conditions.

Starting from the canonical Hamiltonian of the original second-class system
we construct a first-class Hamiltonian with respect to the first-class
subset in two steps \cite{vyt0}. First, we construct the first-class
Hamiltonian with respect to the constraint $G_{1}\approx 0$%
\begin{eqnarray}
H_{GU}^{1}&=&H_{c}-C_{1}\left[ G_{1},H_{c}\right]+\frac{1}{2}C_{1}C_{1}%
\left[G_{1}\left[G_{1},H_{c}\right] \right] -\cdots \nonumber\\
&=&H_{c}+\int d^{2}x\left[ \left(- p_{0}+\partial _{i}\pi ^{i}\right) \left(
\partial _{k}A^{k}+B_{0}\right)\right.\nonumber\\
&&+\frac{1}{m^{2}}\left(- p_{0}+\partial _{i}\pi ^{i}\right) \partial _{k}\partial
^{k}\left( p_{0}+\frac{1}{2b}\varepsilon _{0lm}\partial ^{l}B^{m}\right)\nonumber\\
&&\left.+\frac{1}{2m^{2}}\left(-p_{0}+\partial _{i}\pi ^{i}\right) \partial
_{k}\partial ^{k}\left(-p_{0}+\partial _{j}\pi ^{j}\right) \right] ,
\label{y18}
\end{eqnarray}%
and then, with this at hand, we obtain the first-class Hamiltonian with
respect to the constraint $G_{2}\approx 0$%
\begin{eqnarray}
&&H_{GU} \nonumber\\
&&=H_{GU}^{1}-C_{2}\left[ G_{2},H_{GU}^{1}\right] +\frac{1}{2}C_{2}C_{2}\left[ G_{2}\left[ G_{2},H_{GU}^{1}\right] \right] -\cdots\nonumber\\
&&=H_{GU}^{1}-\int d^{2}x\left[ \left( \partial _{i}A^{i}+B_{0}\right)\partial ^{j}\left( \pi _{j}+\frac{1}{2b}\varepsilon _{0jk}B^{k}\right)\right] .  \label{y20}
\end{eqnarray}%
The Hamiltonian gauge algebra relations are given by
\begin{equation}
[G_{1},H_{GU}] =[G_{2},H_{GU}] =0.
\end{equation}
The equations of motion are
\begin{eqnarray}
\dot{A}_{0} &=&-\frac{1}{m}\partial _{i}\left[ mA^{i}+\frac{1}{m}\partial
^{i}\left( p_{0}+\frac{1}{2b}\varepsilon _{0jk}\partial ^{j}B^{k}\right) %
\right] , \label{r12}\\
\dot{A}_{i} &=&B_{i}-\frac{1}{m^{2}}\partial _{i}\Lambda ^{1}, \label{r13}\\
\dot{p}^{0} &=&-\frac{a}{2}\partial _{i}\left( B^{i}-\partial
^{i}A^{0}\right) +\frac{3}{4b}\varepsilon ^{0ij}\partial _{k}\partial
^{k}\partial _{i}A_{j}+\frac{1}{2}\partial _{i}p^{i}-m^{2}A^{0}+\Lambda ^{1},
\label{r13.1}\\
\dot{p}^{i} &=&a\partial _{j}\partial ^{\lbrack j}A^{i]}-\frac{1}{2b}%
\varepsilon ^{0ij}\partial _{k}\partial ^{k}B_{j}+\frac{1}{b}\varepsilon
^{0ij}\partial _{k}\partial ^{k}\partial _{j}A_{0} \label{r14}\\
&&-m\left[ mA^{i}+\frac{1}{m}\partial ^{i}\left( p_{0}+\frac{1}{2b}%
\varepsilon _{0jk}\partial ^{j}B^{k}\right) \right]-\frac{1}{2b}\varepsilon
^{0ij}\partial _{j}\Lambda ^{2}, \label{r15}\\
\dot{B}_{0} &=&-\Lambda ^{2}+\frac{1}{m^{2}}\partial _{k}\partial
^{k}\Lambda ^{1}, \label{r16}\\
\dot{B}_{i} &=&-ab\varepsilon _{0ij}\left( B^{j}-\partial ^{j}A^{0}\right) -%
\frac{1}{2}\partial _{k}\partial ^{k}A_{i}-\frac{1}{2}\partial _{i}B_{0} \label{r17}-b\varepsilon _{0ij}p^{j}\nonumber\\
&&-\partial _{i}\partial ^{j}\left[ mA_{j}+\frac{1}{%
m}\partial _{j}\left( p_{0}+\frac{1}{2b}\varepsilon _{0kl}\partial
^{k}B^{l}\right) \right] , \label{r18}\\
\dot{\pi}_{0} &=&\frac{1}{2b}\varepsilon _{0ij}\partial ^{i}B^{j}.\label{r19}
\end{eqnarray}
where $\Lambda^{1}$ and $\Lambda^{2}$ are some arbitrary functions. Under the gauge fixing conditions \begin{equation}
B_{0}+\partial_{i}A^{i}\approx 0,\qquad p_{0}+\frac{1}{2b}\varepsilon_{0ij}\partial^{i}B^{j}\approx 0,
\end{equation}
($\Lambda^{1}=0$ and $\Lambda^{2}=\partial_{i}B^{i}$) the equations (\ref{r12})--(\ref{r19}) return to the
equations of motion for the MECS-Proca model.

The number of physical degrees of freedom of the dynamical system with the phase space locally parameterized by $\left\{ A_{\mu },\ B_{\mu },\ p^{\mu },\ \pi
^{\mu }\right\} $, subject to the second-class constraints (\ref{y8.1}) and
first-class constraints (\ref{y15.1}) and (\ref{y15.3}) is equal to%
\begin{eqnarray}
\bar{\mathcal{N}}_{GU} &=&( 12\ \rm{canonical\ variables}-2\rm{ scc}%
-2\times 2\ \rm{fcc}) /2  \nonumber \\
&=&3=\bar{\mathcal{N}}_{O}.  \label{dofguemcs}
\end{eqnarray}

\subsection{St\"{u}ckelberg coupling}

Based on the equivalence between the first-class system and the original
second-class theory, we replace the Hamiltonian path integral of the
MECS-Proca model with that of the first-class system. The Hamiltonian path
integral of the first-class system constructed in the above reads as%
\begin{eqnarray}
Z &=&\int \mathcal{D}\left( A_{\mu },B_{\mu },p^{\mu },\pi ^{\mu },\lambda
^{( 1) },\lambda ^{( 2) }\right) \mu \left( [A_{\mu }],[B_{\mu }]\right) \nonumber\\
&&\times\delta \left[ \pi _{i}+\frac{1}{2b}\varepsilon _{0ij}\left(B^{j}-\partial ^{j}A^{0}\right)\right]\mathrm{det}^{1/2}\left(\frac{1}{b} \varepsilon
_{0ij}\delta( x-y) \right)\nonumber\\
&&\times\exp \left\{ \mathrm{i}\int
d^{3}x\left[ \left( \partial _{0}A_{\mu}\right) p^{\mu}+\left( \partial _{0}B_{\mu}\right)\pi ^{\mu}-\mathcal{H}_{GU}\right.\right.\nonumber\\
&&\left.\left. -\frac{1}{m^{2}}\lambda ^{(1) }\left( \partial
_{i}p^{i}-m^{2}A_{0}+\partial _{k}\partial ^{k}\pi _{0}\right)-\lambda ^{(2) }\left( -\pi _{0}+\frac{1}{2b}\varepsilon _{0ij}\partial
^{i}A^{j}\right) \right] \right\} ,  \label{y21}
\end{eqnarray}
where the integration measure `$\mu \left( [ A_{\mu }],[B_{\mu
}]\right) $' includes some suitable canonical gauge conditions. Performing partial integration over the momenta $\pi _{i}$ in the path
integral, we get to the argument of the exponential in the form%
\begin{eqnarray}
S_{GU} &=&\int d^{3}x\left\{ \left( \partial _{0}A_{\mu}\right) p^{\mu}+\left( \partial _{0}B_{0}\right) \pi
^{0}-\frac{1}{2b}\left( \partial _{0}B_{i}\right) \varepsilon ^{0ij}\left(
B_{j}-\partial _{j}A_{0}\right)\right.\nonumber\\
&&\nonumber\\
&& -\frac{a}{4}\partial _{[
i}A_{j]}\partial ^{[ i}A^{j]}-\frac{a}{2}\left( B_{i}-\partial _{i}A_{0}\right) \left( B^{i}-\partial
^{i}A^{0}\right)\nonumber\\
&& +\frac{1}{2b}\varepsilon _{0ij}\left( \partial _{k}\partial
^{k}A^{0}\right) \partial ^{i}A^{j}+\frac{1}{2b}\varepsilon _{i0j}\left( \partial _{k}\partial
^{k}A^{i}\right) \left(B^{j}-\partial ^{j}A^{0}\right)\nonumber\\
&&-\frac{1}{2}\left[ mA_{i}+\frac{1}{m}\partial _{i}\left( p_{0}+%
\frac{1}{2b}\varepsilon _{0jk}\partial ^{j}B^{k}\right) \right]\nonumber\\
&&\times\left[ mA^{i}+\frac{1}{m}\partial ^{i}\left( p_{0}+\frac{1}{2b}\varepsilon
_{0ln}\partial ^{l}B^{n}\right) \right]-p^{i}B_{i}-\frac{m^{2}}{2}A_{0}A^{0}\nonumber\\
&&+\frac{1}{2b}B^{0}\varepsilon
_{0jk}\partial ^{j}B^{k}-\frac{1}{m^{2}}\lambda ^{(1) }\left( \partial
_{i}p^{i}-m^{2}A_{0}+\partial _{k}\partial ^{k}\pi _{0}\right)\nonumber\\
&&\left.  -\lambda ^{(2) }\left( -\pi _{0}+\frac{1}{2b}\varepsilon _{0ij}\partial
^{i}A^{j}\right) \right\} .  \label{y22}
\end{eqnarray}
Integration over $p^{i}$ leads to a $\delta $ function of the form
\begin{equation}
\delta \left( \partial _{0}A_{i}-B_{i}+\frac{1}{m^{2}}\partial _{i}\lambda ^{(1) }\right),
\end{equation}%
which permits calculation of the integral over $B_{i}$. Performing partial
integration over Lagrange multiplier $\lambda ^{(2) }$ and
$\pi _{0}$, the argument of the exponential becomes%
\begin{eqnarray}
S_{GU} &=&\int d^{3}x\left\{ \left( \partial _{0}A_{0}\right) \left( p^{0}+\frac{%
1}{2b}\varepsilon ^{0ij}\partial _{i}\partial _{0}A_{j}\right)-\frac{a}{4}\partial _{[ i}A_{j]}\partial ^{[ i}A^{j]}\right.\nonumber\\
&&-\frac{a}{2}\left[ \partial _{0}A_{i}-\partial _{i}\left( A_{0}-\frac{1}{m^{2}}\lambda ^{(1) }\right) \right]\left[ \partial ^{0}A^{i}-\partial ^{i}\left(
A^{0}-\frac{1}{m^{2}}\lambda ^{(1) }\right) \right]  \nonumber\\
&&+\frac{1}{2b}\varepsilon _{0ij}\partial _{\lambda }\partial ^{\lambda
}\left( A^{0}-\frac{1}{m^{2}}\lambda ^{(1) }\right) \partial ^{i}A^{j}+\frac{1%
}{2b}\varepsilon _{i0j}\left( \partial _{\lambda }\partial ^{\lambda
}A^{i}\right) \partial ^{0}A^{j}\nonumber\\
&&+\frac{1}{2b}\varepsilon _{ij0}\left( \partial _{\lambda }\partial
^{\lambda }A^{i}\right) \partial ^{j}\left( A^{0}-\frac{1}{m^{2}}\lambda ^{(1)
}\right)\nonumber\\
&&-\frac{1}{2}\left[ mA_{i}+\frac{1}{m}\partial _{i}\left( p_{0}+\frac{1}{2b}\varepsilon _{0jk}\partial ^{j}\partial ^{0}A^{k}\right) \right]\nonumber\\
&&\left.\times\left[ mA^{i}+\frac{1}{m}\partial ^{i}\left( p_{0}+\frac{1}{2b}%
\varepsilon _{0ln}\partial ^{l}\partial ^{0}A^{n}\right) \right]-\frac{m^{2}}{2}A_{0}A^{0}+\lambda ^{(1) }A_{0}\right\} .   \label{y23}
\end{eqnarray}
Making the notations%
\begin{equation}
\varphi=-\frac{1}{m}\left(p^{0}+\frac{1}{2b}\varepsilon ^{0ij}\partial _{i}\partial _{0}A_{j}\right),\quad \bar{A}_{0}=A_{0}-\frac{1}{m^{2}}\lambda ^{( 1) },
\label{y24}
\end{equation}%
and integrating over Lagrange multiplier $\lambda ^{( 1) }$, the
argument of the exponential from the Hamiltonian path integral takes a manifestly Lorentz-covariant form%
\begin{eqnarray}
S_{GU} &=&\int d^{3}x\left[ -\frac{a}{4}\partial _{[ \mu }\bar{A}_{\nu
]}\partial ^{[ \mu }\bar{A}^{\nu ]}+\frac{1}{2b}\varepsilon _{\mu \nu
\rho }\left( \partial _{\lambda }\partial ^{\lambda }\bar{A}^{\mu }\right)
\partial ^{\nu }\bar{A}^{\rho }\right.\nonumber\\
&& \left.-\frac{1}{2}\left( \partial _{\mu }\varphi -m\bar{A}_{\mu }\right)
\left( \partial ^{\mu }\varphi -m\bar{A}^{\mu }\right) \right],  \label{y26}
\end{eqnarray}%
where $\bar{A}_{\mu }=\left\{ \bar{A}_{0},A_{i}\right\} $, and
describes a St\"{u}ckelberg coupling between the scalar
field $\varphi $ and the $1$-form $\bar{A}_{\mu }$. It is obvious that (\ref{y26}) is a higher derivative extension of the result obtained in the previous section (a higher derivative extension involving the CS term). Similar to MCS-Proca model, we find that to St\"{u}ckelberg scalar corresponds a combination of original fields $A_{i}$ and momentum $p^{0}$.

The canonical analysis of the model described by the Lagrangian action (\ref{y26}) displays the constraints (the phase space is locally parameterized by $\left\{ A_{\mu },p^{\mu
},B_{\mu },\pi ^{\mu },\varphi ,p\right\} $)%
\begin{eqnarray}
\chi _{i}&\equiv& \pi _{i}+\frac{1}{2b}\varepsilon _{0ij}\left( B^{j}-\partial
^{j}A^{0}\right) \approx 0  \label{r20}\\
G_{1} &\equiv &\pi _{0}-\frac{1}{2b}\varepsilon _{0ij}\partial
^{i}A^{j}\approx 0,  \label{r21} \\
G_{2} &\equiv &-p_{0}+\partial _{i}\pi ^{i}\approx 0,  \label{r22} \\
G_{3} &\equiv &-\partial _{i}p^{i}+mp-\frac{1}{2b}\varepsilon _{0ij}\partial
_{k}\partial ^{k}\partial ^{i}A^{j}\approx 0,  \label{r23}
\end{eqnarray}%
and the Hamiltonian%
\begin{eqnarray}
H &=&\int d^{2}x\left[ \frac{a}{4}\partial _{\lbrack i}A_{j]}\partial
^{\lbrack i}A^{j]}+\frac{a}{2}\left( B_{i}-\partial _{i}A_{0}\right) \left(
B^{i}-\partial ^{i}A^{0}\right) \right.  \nonumber \\
&&-\frac{1}{2b}\varepsilon _{0ij}\left( \partial _{k}\partial
^{k}A^{0}\right) \partial ^{i}A^{j}-\frac{1}{2b}\varepsilon _{i0j}\left(
\partial _{k}\partial ^{k}A^{i}\right) B^{j}-\frac{1}{2b}\varepsilon _{ij0}\left( \partial _{k}\partial
^{k}A^{i}\right) \partial ^{j}A^{0}  \nonumber \\
&&\left. -p^{\mu }B_{\mu }-\frac{1}{2}p^{2}+mA^{0}p+\frac{1}{2}\left( \partial _{i}\varphi -mA_{i}\right)
\left( \partial ^{i}\varphi -mA^{i}\right) \right] .  \label{r24}
\end{eqnarray}%
The constraints (\ref{r20}) are second-class and the other three constraints are first-class. In order to recover the MECS-Proca model we chose the gauge conditions%
\begin{equation}
C^{1} \equiv \varphi \approx 0, \quad C^{2} \equiv A_{0}\approx 0, \quad C^{3} \equiv B_{0}\approx 0\label{r26.1}
\end{equation}%
such that $\left\{
G_{\Delta},C^{\Delta'}\right\} _{\Delta,\Delta' =\overline{1,3}}$ form a second-class constraints
set and the Hamiltonian path integral is convergent. The Hamiltonian path integral of the gauge system (\ref{y26}) is given by
\begin{eqnarray}
Z &=&\int \mathcal{D}\left( A_{\mu },p^{\mu },B_{\mu },\pi ^{\mu },\varphi
,p\right) \delta \left( \chi _{i}\right) \delta \left( G_{\Delta}\right) \delta
\left( C^{\Delta'}\right)   \nonumber \\
&&\times \exp \left\{ \mathrm{i}\int d^{3}x\left[ \left( \partial _{0}A_{\mu
}\right) p^{\mu }+\left( \partial _{0}B_{\mu }\right) \pi ^{\mu }+\left(
\partial _{0}\varphi \right) p-\mathcal{H}\right] \right\} .  \label{r27}
\end{eqnarray}%
We integrate over the momenta $\left\{ \pi _{i},\pi _{0},p_{0}\right\} $ and
fields $\{\varphi ,A_{0}\}$ and represent $\delta \left( -\partial
_{i}p^{i}+mp-\frac{1}{2b}\varepsilon _{0ij}\partial _{k}\partial
^{k}\partial ^{i}A^{j}\right) $ in the form of integral functional
\begin{equation}
\int \mathcal{D}\lambda \exp \left\{ -\mathrm{i}\int d^{3}x\lambda \left(
-\partial _{i}p^{i}+mp-\frac{1}{2b}\varepsilon _{0ij}\partial _{k}\partial
^{k}\partial ^{i}A^{j}\right) \right\} .  \label{r28}
\end{equation}%
The path integral takes the form%
\begin{eqnarray}
Z &=&\int \mathcal{D}\left( A_{i},p^{i},B_{\mu },p,\lambda \right) \delta
\left( C^{3}\right) \exp \left\{ \mathrm{i}\int d^{3}x\left[ \left( \partial
_{0}A_{i}\right) p^{i}\right. \right.   \nonumber \\
&&-\frac{1}{2b}\varepsilon ^{0ij}\left( \partial
_{0}B_{i}\right) B_{j}+\frac{1}{2b}\varepsilon ^{0ij}\left( \partial ^{0}B_{0}\right) \partial
_{i}A_{j}-\frac{a}{4}\partial _{\lbrack i}A_{j]}\partial ^{\lbrack i}A^{j]}-
\frac{a}{2}B_{i}B^{i}\nonumber \\
&&+\frac{1}{2b}\varepsilon _{i0j}\left( \partial
_{k}\partial ^{k}A^{i}\right) B^{j}+\frac{1}{2b}\varepsilon _{0ij}B^{0}\partial ^{i}B^{j}-p_{i}B^{i}+\frac{1}{%
2}p^{2}-\frac{m^{2}}{2}A_{i}A^{i}  \nonumber \\
&&\left. \left. -\lambda \left( -\partial _{i}p^{i}+mp-\frac{1}{2b}%
\varepsilon _{0ij}\partial _{k}\partial ^{k}\partial ^{i}A^{j}\right) \right]
\right\} .  \label{r29}
\end{eqnarray}%
Integration over $p^{i}$ leads to a $\delta $ function of the form
\begin{equation}
\delta \left( \partial _{0}A_{i}-B_{i}-\partial _{i}\lambda \right) ,
\label{r30}
\end{equation}%
which permits calculation of the integral over $B_{i}$. After integration
over the momentum $p$ and field $B_{0}$, the path integral read as
\begin{eqnarray}
Z &=&\int \mathcal{D}\left( A_{i},\lambda \right) \exp \left\{ \mathrm{i}\int d^{3}x\left( -\frac{a}{4}\partial
_{\lbrack i}A_{j]}\partial ^{\lbrack i}A^{j]}\right. \right.   \nonumber \\
&&-\frac{a}{2}\left( \partial _{0}A_{i}-\partial _{i}\lambda \right) \left(
\partial ^{0}A^{i}-\partial ^{i}\lambda \right)+\frac{1}{2b}\varepsilon
^{i0j}\left( \partial _{\mu }\partial ^{\mu }A_{i}\right) \partial _{0}A_{j}\nonumber \\
&&\left. \left.  +\frac{1}{2b}\varepsilon ^{0ij}\left( \partial _{\mu
}\partial ^{\mu }\lambda \right)\partial _{i}A_{j}+\frac{1}{2b}\varepsilon
^{ij0}\left( \partial _{\mu }\partial ^{\mu }A_{i}\right) \partial
_{j}\lambda -\frac{m^{2}}{2}A_{\mu }A^{\mu }\right) \right\} .  \label{r31}
\end{eqnarray}%
Making the notation $A_{0}=\lambda$ the argument of the exponential from the Hamiltonian path integral is
exactly the MECS-Proca Lagrangian%
\begin{eqnarray}
Z &=&\int \mathcal{D} A_{\mu }
\exp \left\{ \mathrm{i}\int d^{3}x\left( -\frac{a}{4}\partial _{\lbrack \mu
}A_{\nu ]}\partial ^{\lbrack \mu }A^{\nu ]}\right. \right.   \nonumber \\
&&\left. \left. +\frac{1}{2b}\varepsilon ^{\mu \nu \rho }\left( \partial
_{\lambda }\partial ^{\lambda }A_{\mu }\right) \partial _{\nu }A_{\rho }-%
\frac{m^{2}}{2}A_{\mu }A^{\mu }\right) \right\} .
\end{eqnarray}
\subsection{Chern-Simons coupling}

In the sequel we show that the MECS-Proca model may be related to another
first-class theory. Starting from the GU system constructed in the above,
subject to the second-class constraints (\ref{y8.1}), the first-class
constraints (\ref{y15.1}) and (\ref{y15.3}) and whose evolution is governed
by the first-class Hamiltonian (\ref{y20}), we consider the following fields/momenta
combinations%
\begin{eqnarray}
&&\mathcal{F}_{0} \equiv A_{0},\quad \mathcal{F}_{i}\equiv A_{i}+\frac{1}{m^{2}}\partial
_{i}\left( p_{0}-\partial_{j}\pi^{j}\right) ,
\label{y28.2} \\
&&\mathcal{P}_{i} \equiv p_{i}-\frac{1}{2b}\varepsilon _{0ij}\partial
_{k}\partial ^{k}A^{j}-\frac{1}{2b}\varepsilon _{0ij}\partial
^{j}B^{0},\quad \mathcal{B}_{i}\equiv B_{i},  \label{y28.3}
\end{eqnarray}%
which are in (strong) involution with first-class constraints $G_{a}\approx
0 $%
\begin{equation}
\left[ \mathcal{F}_{0},G_{a}\right] =\left[ \mathcal{F}_{i},G_{a}\right] =%
\left[ \mathcal{P}_{i},G_{a}\right] =\left[ \mathcal{B}_{i},G_{a}\right] =0,
\label{y29}
\end{equation}%
and, moreover, $\mathcal{F}_{\mu }\equiv \left\{ \mathcal{F}_{0},\mathcal{F}%
_{i}\right\} $ is divergenceless on the surface $\chi _{i}^{\left( 1\right)}\approx0$%
\begin{equation}
\partial ^{\mu }\mathcal{F}_{\mu }=\mathcal{O}\left(\chi _{i}^{\left( 1\right)}\right).  \label{y30}
\end{equation}%
Similarly to the case of the MCS-Proca model, the first-class Hamiltonian (%
\ref{y20}) can be written in terms of these quantities%
\begin{eqnarray}
H_{GU} &=&\int d^{2}x\left[ \frac{a}{4}\partial _{[ i}\mathcal{F}%
_{j]}\partial ^{[ i}\mathcal{F}^{j]} +\frac{a}{2}\left( \mathcal{B}%
_{i}-\partial _{i}\mathcal{F}_{0}\right)\left( \mathcal{B}^{i}-\partial ^{i}\mathcal{F}^{0}\right)\right.\nonumber\\
&&\nonumber\\
&&-\frac{1}{2b}\varepsilon _{0ij}\left( \partial _{k}\partial ^{k}\mathcal{F}%
^{0}\right) \partial ^{i}\mathcal{F}^{j}-\frac{1}{2b}\varepsilon _{ij0}\left( \partial _{k}\partial ^{k}\mathcal{F}%
^{i}\right) \partial ^{j}\mathcal{F}^{0}\nonumber\\
&&\left.+\frac{m^{2}}{2}\mathcal{F}_{i}\mathcal{F}^{i}+\frac{m^{2}}{2}\mathcal{F}_{0}\mathcal{F}^{0}+\mathcal{B}^{i}\mathcal{P}%
_{i}-\left(\partial^{i}\mathcal{F}_{i}\right)\partial^{j}\chi _{j}^{( 1) }\right] .  \label{y27}
\end{eqnarray}%
Enlarging the phase space by adding the bosonic pairs $\left\{ V^{\mu },P_{\mu
}\right\}$, the solution to the Eq. (\ref{y30}) takes the form
\begin{equation}
\mathcal{F}_{\mu }=-\frac{1}{m}\varepsilon _{\mu \nu \rho }\partial ^{\nu }V^{\rho }.
\label{y31}
\end{equation}%
When we replace the solution (\ref{y31}) in (\ref{y15.1}), the constraint takes the form
\begin{equation}
\frac{1}{m^{2}}\left(\partial _{i}p^{i}+m\varepsilon _{0ij}\partial ^{i}V^{j}+\partial
_{k}\partial ^{k}\pi _{0}\right)\approx 0,  \label{y32}
\end{equation}%
and remains first-class. Computing the Poisson bracket among the quantity $%
\partial _{i}p_{0}$ and first-class constraint (\ref{y15.1}) and the Poisson
bracket between $P_{i}$ and (\ref{y32}), we obtain that these two quantities
are correlated through the relation%
\begin{equation}
\partial _{i}p_{0}=m\varepsilon _{0ij}P^{j}.  \label{y34}
\end{equation}%
Using the relations (\ref{y31}) and (\ref{y34}), we write the first-class Hamiltonian as
\begin{eqnarray}
H_{GU}^{\prime }&=&\int d^{2}x\left\{ \frac{a}{4}\partial _{\lbrack
i}A_{j]}\partial ^{\lbrack i}A^{j]}\right. \nonumber\\
&&+\frac{a}{2}\left[ B_{i}+\frac{1}{m}\partial
_{i}\left( \varepsilon _{0jk}\partial ^{j}V^{k}\right) \right]\left[ B^{i}+\frac{1}{m}\partial ^{i}\left( \varepsilon ^{0ln}\partial
_{l}V_{n}\right)\right]\nonumber\\
&&+\frac{1}{2b}\varepsilon _{0ij}\partial
_{k}\partial ^{k}\left(\frac{1}{m} \varepsilon ^{0ln}\partial _{l}V_{n}\right)\partial ^{i}A^{j}-\frac{1}{2b}\varepsilon _{i0j}\left( \partial _{k}\partial
^{k}A^{i}\right) B^{j}\nonumber\\
&& +\frac{1}{2b}\varepsilon _{ij0}\left( \partial
_{k}\partial ^{k}A^{i}\right) \partial ^{j}\left(\frac{1}{m}\varepsilon ^{0ln}\partial
_{l}V_{n}\right)+\frac{1}{4}\partial ^{[ i}V^{j]}\partial _{[ i}V_{j]}\nonumber\\
&&+\frac{m^{2}}{2}\left( A_{i}+\frac{1}{m}\varepsilon _{0ij}P^{j}-\frac{1}{m^{2}}\partial _{i}\partial
_{j}\pi ^{j}\right)\left( A^{i}+\frac{1}{m}\varepsilon ^{0il}P_{l}-\frac{1}{m^{2}}\partial
^{i}\partial ^{l}\pi _{l}\right) \nonumber\\
&&-\partial ^{i}\left( A_{i}+\frac{1}{m}\varepsilon _{0ik}P^{k}-\frac{1}{m^{2}}\partial
_{i}\partial _{k}\pi ^{k}\right) \partial ^{j}\left( \pi _{j}+\frac{1}{2b}%
\varepsilon _{0jk}B^{k}\right)
\nonumber\\
&&\left.
-\frac{1}{2b}\varepsilon _{0jk}B^{0}\partial ^{j}B^{k}  +p^{i}B_{i}\right\} .  \label{y35}
\end{eqnarray}
If we count the number of physical degrees of freedom of the system with the
phase space locally parameterized by $\left\{ A_{i},B_{\mu },V^{\mu
},p^{i},\pi ^{\mu },P_{\mu }\right\} $ subject to the second-class
constraints (\ref{y8.1}), first-class constraints (\ref{y15.3}) and (\ref%
{y32}) and whose evolution is governed by the first-class Hamiltonian (\ref%
{y35}), we obtain%
\begin{eqnarray}
\bar{\mathcal{N}}_{GU}^{\prime } &=&(16\ \rm{canonical\ variables}-2%
\rm{ scc}-2\times 2\ \rm{fcc}) /2  \nonumber \\
&=&5\neq \bar{\mathcal{N}}_{GU}.  \label{dofguemcs'}
\end{eqnarray}%
Imposing the first-class constraints
\begin{equation}
-\partial ^{i}P_{i}\approx 0,\qquad P_{0}\approx 0,  \label{y33}
\end{equation}%
the number of physical degrees of freedom is conserved
\begin{eqnarray}
\bar{\mathcal{N}}_{GU}^{\prime } &=&( 16\ \rm{canonical\ variables}-2%
\rm{ scc}-2\times 4\ \rm{fcc}) /2  \nonumber \\
&=&3=\bar{\mathcal{N}}_{GU}.
\end{eqnarray}%
For each first-class theory, derived in the above, we are able to identify a
set of fundamental classical observables such that they are in one-to-one
correspondence and they possess the same Poisson brackets. Since the number of physical degrees of freedom is the same for both theories and the corresponding algebras of classical observables are
isomorphic, the previously exposed procedure preserves the equivalence between the two first-class theories. As a result, the GU and the first-class system remain equivalent
also at the level of the Hamiltonian path integral quantization. This
further implies that the first-class system is completely equivalent with
the MECS-Proca model. Due to this equivalence we can replace the Hamiltonian
path integral of the MECS-Proca model with that one associated with the
first-class system
\begin{eqnarray}
Z^{\prime }&=&\int \mathcal{D}\left( A_{i},B_{\mu },V^{\mu },p^{i},\pi
^{\mu },P_{\mu },\lambda's\right)  \mu \left(
[ A_{i}],[B_{\mu }],[V^{\mu }]\right)\nonumber\\
&&\times\delta \left[ \pi _{i}+\frac{1}{2b}\varepsilon _{0ij}\left(B^{j}+\frac{1}{m}\varepsilon ^{0kl}\partial ^{j}\partial
_{k}V_{l}\right) \right]\mathrm{det}^{1/2}\left(\frac{1}{b} \varepsilon _{0ij}\delta ( x-y)\right)\nonumber\\
&&\times\exp\left\{\mathrm{i}\int d^{3}x\left[ \left( \partial
_{0}A_{i}\right) p^{i}+\left( \partial _{0}B_{\mu}\right) \pi ^{\mu}+\left( \partial _{0}V^{\mu}\right)
P_{\mu}-\mathcal{H}_{GU}^{\prime }\right.\right.\nonumber\\
&&-\frac{1}{m^{2}}\lambda
^{( 1) }\left( \partial _{i}p^{i}+m\varepsilon _{0ij}\partial
^{i}V^{j}+\partial _{k}\partial ^{k}\pi _{0}\right)\nonumber\\
&&\left.\left.-\lambda ^{( 2) }\left( -\pi _{0}+\frac{1}{2b}%
\varepsilon _{0ij}\partial ^{i}A^{j}\right)+\lambda ^{(3) }\partial ^{i}P_{i}-\lambda ^{(4)
}P_{0}\right] \right\} .
\label{y36}
\end{eqnarray}
After a partial integration over the momenta $\pi _{i}$ in the path integral,
the argument of the exponential read as%
\begin{eqnarray}
S_{GU}^{\prime } &=&\int d^{3}x\left\{ \left( \partial _{0}A_{i}\right)
p^{i}+\left( \partial _{0}B_{0}\right) \pi ^{0}+\left( \partial
_{0}V^{\mu}\right) P_{\mu}\right. \nonumber\\
&&+\frac{1}{2b}\left( \partial _{0}B_{i}\right) \varepsilon ^{0ij}\left[
-B_{j}-\partial _{j}\left( \frac{1}{m}\varepsilon ^{0kl}\partial
_{k}V_{l}\right) \right]-\frac{a}{4}\partial _{[i}A_{j]}\partial
^{[i}A^{j]} \nonumber\\
&& -\frac{a}{2}\left[ B_{i}+\partial _{i}\left( \frac{1}{m}\varepsilon _{0jk}\partial
^{j}V^{k}\right) \right]\left[ B^{i}+\partial ^{i}\left( \frac{1}{m}\varepsilon
^{0ln}\partial _{l}V_{n}\right) \right] \nonumber\\
&& -\frac{1}{2b}\varepsilon _{0ij} \partial _{k}\partial ^{k}\left(\frac{1}{m}
\varepsilon ^{0ln}\partial _{l}V_{n}\right) \partial ^{i}A^{j}+\frac{1}{2b}\varepsilon _{i0j}\left( \partial _{k}\partial ^{k}A^{i}\right)
B^{j}\nonumber\\
&&-\frac{1}{2b}\varepsilon _{ij0}\left( \partial _{k}\partial
^{k}A^{i}\right)\partial ^{j}\left( \frac{1}{m}\varepsilon ^{0ln}\partial
_{l}V_{n}\right)-\frac{1}{4}\partial ^{\lbrack i}V^{j]}\partial _{\lbrack
i}V_{j]} \nonumber\\
&& -\frac{m^{2}}{2}\left[ A_{i}+\frac{1}{m}\varepsilon _{0ij}P^{j}+\frac{1}{m^{2}}\partial _{i}\left(\frac{1}{2b}\varepsilon
_{0jk}\partial ^{j}B^{k}\right)\right]\nonumber\\
&&\times\left[ A^{i}+\frac{1}{m}\varepsilon^{0il}P_{l}+\frac{1}{m^{2}}\partial _{i}\left(\frac{1}{2b}\varepsilon _{0ln}\partial ^{l}B^{n}%
\right)\right]\nonumber\\
&& +\frac{1}{2b}\varepsilon _{0jk}B^{0}\partial ^{j}B^{k}-p^{i}B_{i}-\frac{1}{m^{2}}\lambda
^{( 1) }\left( \partial _{i}p^{i}+m\varepsilon _{0ij}\partial
^{i}V^{j}+\partial _{k}\partial ^{k}\pi _{0}\right) \nonumber\\
&& \left. -\lambda ^{(2) }\left( -\pi _{0}+\frac{1}{2b}%
\varepsilon _{0ij}\partial ^{i}A^{j}\right)+\lambda ^{(3) }\partial ^{i}P_{i} -\lambda ^{(4)
}P_{0}\right\} .  \label{y37}
\end{eqnarray}
Integration over $p^{i}$ leads to a $\delta $ function of the form
\begin{equation}
\delta \left( \partial _{0}A_{i}-B_{i}+\frac{1}{m^{2}}\partial _{i}\lambda ^{(1) }\right) ,
\end{equation}%
which permits calculation of the integral over $B_{i}$. Performing partial
integration over the field $V_{0}$, momenta $\left\{\pi _{0},P_{0},P_{i}\right\}$
and Lagrange multipliers $\left\{\lambda ^{(2)},\lambda ^{(4) }\right\}$, the argument of the exponential from the Hamiltonian path
integral reads as%
\begin{eqnarray}
S_{GU}^{\prime } &=&\int d^{3}x\left\{ -\frac{a}{4}\partial _{[i}A_{j]}\partial ^{[ i}A^{j]}-\frac{a}{2}\left[ \partial
_{0}A_{i}+\partial _{i}\left(\frac{1}{m^{2}}\lambda ^{(1) }+\frac{1}{m}\varepsilon_{0jk}\partial ^{j}V^{k}\right) \right] \right. \nonumber  \\
&& \times\left[ \partial ^{0}A^{i}+\partial ^{i}\left( \frac{1}{m^{2}}\lambda ^{(1) }+\frac{1}{m}\varepsilon ^{0ln}\partial _{l}V_{n}\right) \right]\nonumber  \\
&&-\frac{1}{2b}\varepsilon _{0ij}\partial _{\lambda }\partial ^{\lambda
}\left(\frac{1}{m^{2}} \lambda ^{(1) }+\frac{1}{m}\varepsilon ^{0kl}\partial
_{k}V_{l}\right) \partial ^{i}A^{j}+\frac{1}{2b}\varepsilon _{i0j}\left(\partial _{\lambda }\partial ^{\lambda }A^{i}\right)\partial ^{0}A^{j}\nonumber  \\
&& -\frac{1}{2b}\varepsilon _{ij0}\left(\partial _{\lambda }\partial ^{\lambda }A^{i}\right)\partial ^{j}\left(\frac{1}{m^{2}} \lambda ^{(1)
}+\frac{1}{m}\varepsilon ^{0kl}\partial _{k}V_{l}\right)\nonumber  \\
&&+\frac{1}{4}\partial _{[ i}V_{j]}\partial ^{[ i}V^{j]}+\frac{%
1}{2}\left( \partial _{0}V_{i}-\partial _{i}\lambda ^{(3)
}\right) \left( \partial ^{0}V^{i}-\partial ^{i}\lambda ^{(3)
}\right) \nonumber\\
&&-m\varepsilon _{0ij}\left( \frac{1}{m^{2}}\lambda
^{(1) }+\frac{1}{m}\varepsilon ^{0kl}\partial _{k}V_{l}\right) \left( \partial ^{i}V^{j}\right)\nonumber\\
&&\left.+m\varepsilon _{i0j}A^{i}\left( \partial ^{0}V^{j}-\partial ^{j}\lambda ^{(3) }\right) \right\} .
\label{y39}
\end{eqnarray}
Using the notations%
\begin{equation}
\bar{A}_{0}=-\left(\frac{1}{m^{2}}\lambda ^{( 1) }+\frac{1}{m}\varepsilon _{0jk}\partial ^{j}V^{k}\right),\qquad \bar{V}_{0}=\lambda ^{(3) },  \label{y40}
\end{equation}%
the argument of the exponential from the Hamiltonian path integral takes a manifestly Lorentz-covariant form%
\begin{eqnarray}
S_{GU}^{\prime } &=&\int d^{3}x\left[ -\frac{a}{4}\partial _{[ \mu }%
\bar{A}_{\nu ]}\partial ^{[ \mu }\bar{A}^{\nu ]}+\frac{1}{2b}%
\varepsilon _{\mu \nu \rho }\left( \partial _{\lambda }\partial ^{\lambda }%
\bar{A}^{\mu }\right) \partial ^{\nu }\bar{A}^{\rho }\right.  \nonumber \\
&&\left.+\frac{1}{4}\partial _{[\mu }\bar{V}_{\nu ]}\partial
^{[ \mu }\bar{V}^{\nu ]}+m\varepsilon _{\mu \nu \rho }\bar{A}^{\mu } \partial
^{\nu }\bar{V}^{\rho } \right] ,
\end{eqnarray}%
where $\bar{A}_{\mu }=\left\{ \bar{A}_{0},A_{i}\right\} $ and $\bar{V%
}_{\mu }=\left\{ \bar{V}_{0},V_{i}\right\} $. The above functional
describes a CS coupling between the $1$-form $%
\bar{A}_{\mu }$ and the $1$-form $\bar{V}_{\mu }$ and it is a higher derivative extension of the functional (\ref{x19}).

\section{Conclusions\label{sec4}}

In this paper, the MCS-Proca model has been analyzed from the point of view of the Hamiltonian path
integral quantization, in the framework of gauge-unfixing approach. The same quantization procedure was applied to a higher order derivative extension of MCS-Proca model. The first step of this approach is represented by the construction of an equivalent first-class system. In order to construct the equivalent
first-class system with MECS-Proca model, we performed a partial gauge-unfixing (we maintained the second-class constraints (\ref{y8.1})), meanwhile in the case of the MCS-Proca model we accomplished a total gauge-unfixing. Both models did not require extensions of the original phase space in order to construct the equivalent first-class systems. The second step involved the construction of the Hamiltonian path integral corresponding to the equivalent first-class system for each model. The Hamiltonian path integral of the first-class systems took a manifestly Lorentz-covariant form, after integrating out the auxiliary fields and performing some field redefinitions. Starting from the Hamiltonian path integral of the equivalent non-higher derivative first-class system, we arrived to the Lagrangian path integral corresponding to St\"{u}ckelberg coupling between a scalar field and a $1$-form or for an appropriate phase space extensions we identified the Lagrangian path integral for two kinds of $1$-forms with CS coupling (a non-higher order derivative term). The results obtained in the case of MECS-Proca model are higher derivative extensions (involving the CS term) of the results obtained in the case of MCS-Proca model.

\section*{Acknowledgement}
The author wishes to thank E.M. Cioroianu for useful discussions and
comments. I am very grateful to Prof. S. Deser for calling my attention to the Ref. \cite{DT}.


\end{document}